# ON SPACE'S TOPOLOGY

By

AMOS ALTSHULER

Dept. of Mathematics, Ben-Gurion Univ. of the Negev, Beer-Sheva, Israel

The human mind has first to construct forms,
 independently, before we can find them in things.
                                        Albert Einstein

## ABSTRACT

It seems to be a common belief that the space in which we live is a space-time manifold of dimension at least four. In the present article we wish to draw attention to a slightly different possibility - A space- time pseudomanifold (or even a generalized pseudomanifold,of one dimension or another). Basically, a pseudomanifold is in a sense a manifold K with certain irremovable singularities inherited in the topological structure of K, and prior and independent to any metric or differential structure imposed on K. We explain this term for dimension 3, but it holds analogously for higher dimensions as well ([5], [2, Sec.4])).

## 1. INTRODUCTION

The classical assumption is that the space component M(3) of the space-time R(1)×M(3) is a 3-dimensional manifold, and the big problem is to find out **which** 3-manifold it is. This assumption stems from two facts:
A. Our experience tells us that the neighborhood of each point in space which we get our hand on is (topologically) a 3-dimensional ball.
B. It is unconceivable to think of a point anywhere in our space whose neighborhood is not a 3-dimensional ball.

A 3-dimensional manifold (abbreviated: 3-manifold) is a (mathematical) space in which the neighborhood of every point is a 3-ball. Thus the natural conclusion is that the space in which we live is a 3-manifold, and the question we face is to find out **which** manifold it is: a 3-sphere ? a 3-torus ? perhaps some non-orientable 3-manifold?.

(Note: our terminology is the one used in **mathematics**. Thus a 2-sphere is the boundary of the (topological) 3-dimensional (solid) ball, and a 3-sphere is the boundary of the (topological) 4-dimensional (solid) ball.)

However, there exist (mathematical) spaces which almost everywhere look like 3-manifolds, but are not 3-manifolds. They are called 3-pseudomanifolds. A 3-pseudomanifold (abbreviated 3-pm) looks like a 3-manifold everywhere but a finite (or infinite and isolated) set of **singular** points, and those singular points are essential and not removable. Why not consider the possibility that the space component of the space-time we live in is such a 3-pm (or, analogously, in case of considering manifolds of some other dimension n, an n-pseudomanifold )? It may be that certain phenomena can be explained by the existence of singular points, namely by the mere structure of space itself.

The definition of a 3-pseudomanifold is based on the concept of 2-manifold. Thus Section 2 is devoted to a brief description of 2-manifolds. ( For a more detailed treatment see e.g. [4] or [6].) This ennables the description, in Section 3, of a first example of a 3-pm. However, the general treatment of 3-pms is based on **triangulated manifolds**, described in Section 4 . In Section 5, another example of a 3-pm is given and the main theorem concerning 3-pms is stated. Finally, in Section 6, we briefly describe more general pseudomanifolds, namely **generalized pseudomanifolds** which perhaps too should be considered as possible space component in our space-time universe.

## 2. 2- MANIFOLDS

A 2-dimensional manifold (abbreviated: 2-manifold) is a (mathematical) space in which the neighborhood of every point is a (topological) disc. Thus the Euclidean plane $\mathbf{R}^2$ is a 2-manifold, but not a compact one. The disjoint union of two 2-spheres is a compact 2- manifold, but not connected. From now on we assume all the 2-manifolds in the present article to be connected and compact .

The 2-manifolds are sorted into two classes. The **orientable** 2- manifolds are those that , like the 2-sphere and the torus (Figs. 1,2), can be realized in the Euclidean 3-space $\mathbf{R}^3$ ; the **non-orientable** ones are those that cannot be realized in $\mathbf{R}^3$ . Although for the concrete physical construction of the orientable 2- manifolds we need the space $\mathbf{R}^3$, they can be faithfully described in the plane $\mathbf{R}^2$ , by using the method of **identification** (or **gluing**) of edges, as follows.

In Fig.1a the 2-sphere is described as a disc whose boundary is composed of two directed arcs that are assumed to be identified to each other according to the given directions. Doing it physically means bending the disc (Fig. 1b) and gluing the two arcs to each other (Fig. 1c), to yield the 2-sphere $\mathbf{S}^2$.

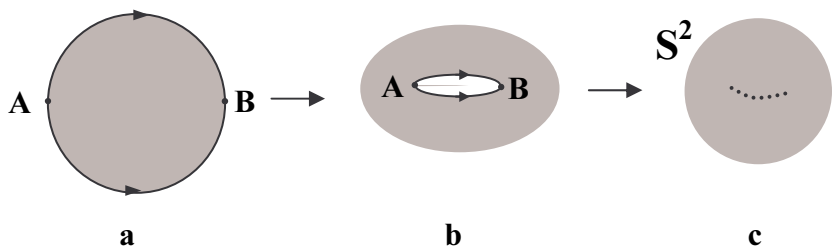

a          b          c

Figure 1: The 2-sphere $\mathbf{S}^2$.

Similarly, the torus can be faithfully described as a rectangle with two identifications (Fig. 2a): the horizontal edges are identified to each other (doing it physically yields a cylinder (Fig.2b)) and the two vertical edges (which in the cylinder become two oriented circles) are identified to each other (the cylinder is bent so that the two circles are glued to each other according to the given directions (Figs. 2c,d)) to yield the torus $\mathbf{T}^2$. The four vertices of the original rectangle are thus identified to form one single point, whose neighborhood, originally seen as four quarters of a disc, is a disc.

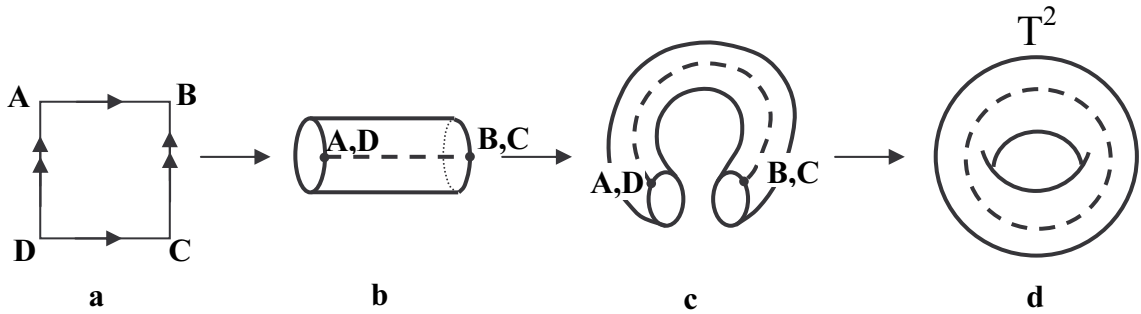

Figure 2: The torus $T^2$

In the same manner, the non-orientable 2- manifolds, though not constructible in $R^3$, can be faithfully described in $R^2$ : the boundary of the disc in Fig. 3a is again divided into two oriented arcs which should be identified according to the given directions, but this abstract identification cannot be followed by an actual gluing in $R^3$ (however it can be done in some higher dimensional Euclidean space). This yields the non-orientable 2-manifold known as the **projective plane $P^2$** . Similarly. the identification of the edges in the rectangle depicted in Fig. 3b, as hinted by the arrows, yields the non-orientable 2-manifold known as the **Klein bottle $K^2$**.

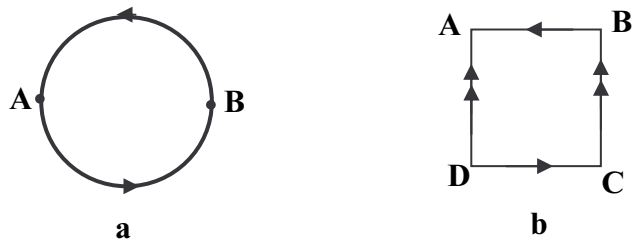

Figure 3; The projective plane $P^2$ and the Klein bottle $K^2$.

Within each of the two classes, the orientable and the non-orientable 2-manifolds, the 2-manifolds are characterized by an integer known as the **Euler characteristic**. A simple method for finding the Euler characteristic $\chi(M)$ of a 2-manifold M is achieved by constructing a "map" on M, which covers the entire of M, such that each "country" in this map is topologically a disc. The countries are called **faces** of the map, the borders in which two countries meet are the **edges** and the edges meet at the **vertices** of the map. Now consider any such map on M, and let V, E and F be the number of vertices, edges and faces resp. in the map. The Euler characteristic $\chi(M)$ of M is defined as $\chi(M) = V-E+F$ and is independent of the particular map chosen on M.

The sphere $S^2$ in Fig. 1 is already endowed with a (very simple) map, composed of one face (the disc), one edge AB and two vertices A, B. Hence $\chi(S^2)=2$. In the projective plane $P^2$ of Fig. 3a we have a similar map composed of one face and one edge. But here there is only one vertex, as the directions of the two arcs cause the two points A, B to be identified with each other and form one single vertex. Thus $\chi(P^2)=1$. In Fig. 11 another map is depicted on the projective plane, this time with six vertices

(of which three are on the "boundary"), 15 edges (three on the "boundary") and 10 faces (all triangles), and again $\chi(\mathbf{P}^2)$=6-15+10=1.

The torus of Fig. 2 is already endowed with a map composed of one face (the rectangle), two edges (AB=CD, AD=BC) and one vertex (A=B=C=D), hence $\chi(\mathbf{T}^2)$=0. In Fig.10 another map on the torus is depicted, with 14 faces (all triangles), 21 edges (3 on the horizontal side of the rectangle, 3 on the vertical side and 15 interior to the rectangle) and 7 vertices, and again $\chi(\mathbf{T}^2)$=7-21+14=0. The Klein bottle $\mathbf{K}^2$ of Fig. 3b, like the torus in **F**ig. 2a, has V=1, E=2, F=1, so its Euler characteristic, like that of the torus, is 0.

For every orientable 2-manifold M, the Euler characteristic $\chi(M)$ is an even integer ≤2, and vice versa: for every even integer k≤2, there is a (topologically) unique orientable 2-manifold M such that $\chi(M)$=k. The **genus** $\gamma(M)$ of an orientable 2-manifold M is defined to be $\gamma(M)=½(2-\chi(M))$. Figure 4 depicts the (orientable) 2-manifolds of genus 0, 1, 2 and 3.

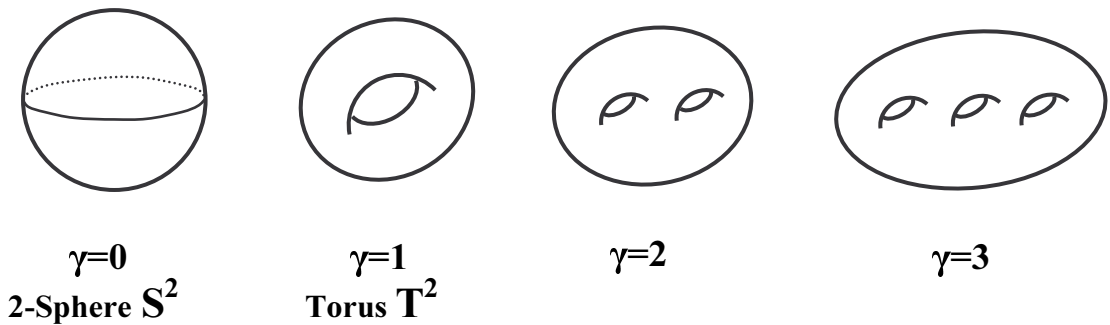

$\gamma$=0  $\gamma$=1  $\gamma$=2  $\gamma$=3
**2-Sphere** $\mathbf{S}^2$    **Torus** $\mathbf{T}^2$

Figure 4: Orientable 2-manifolds of genus 0, 1, 2 and 3.

For a non-orientable 2-manifold M, $\chi(M)$ is an integer ≤1, and vice versa: for every integer k≤1 there is a (topologically) unique non-orientable 2-manifold M such that $\chi(M)$=k. The **connectivity** q(M) of a non-orientable 2-manifold M is defined as q(M)= 2- $\chi(M)$. Thus the connectivity of the projective plane is q($\mathbf{P}^2$)=1 and that of the Klein bottle is q($\mathbf{K}^2$)=2.

## 3. A 3-manifold and a 3-pseudomanifold

A 3-manifold is a (mathematical) space in which the neighborhood of every point is a (topological) 3-ball.
The boundary of the neighborhood of a point A in some space M is called the **link** of A in M and is denoted link(A,M), briefly linkA. The requirement that the neighborhood of a point A be a (topological) 3-ball is equivalent to the requirement that linkA be a (topological) 2-sphere. Thus we get an equivalent definition of a 3-manifold:
  A 3-manifold is a (mathematical) space in which the link of every point is a (topological) 2-sphere.
 If we replace here "2-sphere" by "2-manifold", we get the definition of a **3-dimensional pseudomanifold:**

A **3-dimensional pseudomanifold** (abbreviated 3-pm) **is** is a (mathematical) space in which the link of every point is **a** (topological) 2-manifold.

A point in a 3-pm is **regular** if its link is a 2-sphere, otherwise it is **singular**. If the link of a singular point A is a 2-manifold M, we say that the singularity of A is M.

Let us consider two examples.

Consider the cube with vertices ( ±1, ±1, ±1) and identify all points in the cube of the form (1, y, z) to (-1, y, z), (x, 1, z) to (x, -1, z) and (x, y, 1) to (x, y, -1). That is: each point in a wall (or floor) of the cube is identified with the point in the parallel wall (or ceiling, resp.) directly opposite to it. This identification yields from the cube a 3-manifold as follows.

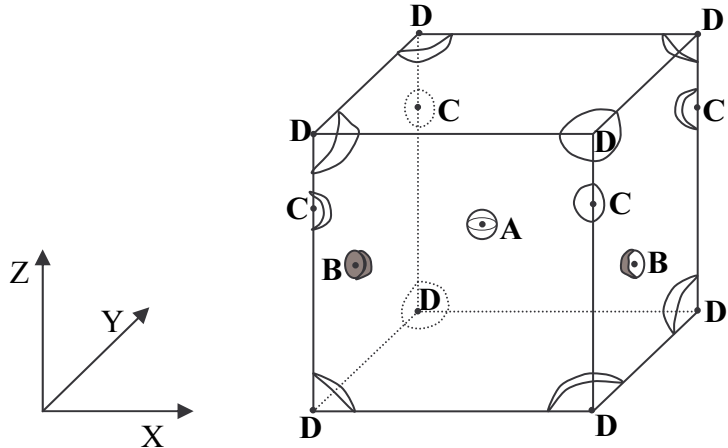

Figure 5:   The 3-manifold $\mathbf{T}^3$

The neighborhood of an interior point (A, in Fig. 5) is obviously a 3-ball. The neighborhood of each point B interior to a wall is made of two half 3-balls which fit together to form a 3-ball. Thus linkB (like linkA) is a 2-sphere $\mathbf{S}^2$. Similarly the neighborhood of a point C interior to an edge of the cube is a 3-ball composed of four quarters of a 3-ball ( a careful examination reveals that these four quarters do indeed fit together to form a 3-ball). All the eight vertices form one single point D by our identification, and the neighborhood of this point is composed of  eight eighths of a 3-ball which fit nicely together to form a 3-ball.
Thus this identification yields from the cube a 3-manifold, known as  the 3-dimensional torus $\mathbf{T}^3$.

Now consider the same cube with a different identification. (See [6,Sec. 10.3].) Points of the form (1, y, z) are identified to (-1, -y, z), (x, 1, z) to (x, -1, -z) and (x, y, 1) to ((-x, y, -1). Thus each wall (or floor) is glued to the opposite wall (or ceiling, resp.), but with a flip: the front and back walls by a flip on a vertical axis in the plane y=0, the right and left walls by a flip on a horizontal axis in the plane z=0 and the floor and ceiling by a flip on a horizontal axis in the plane x=0.

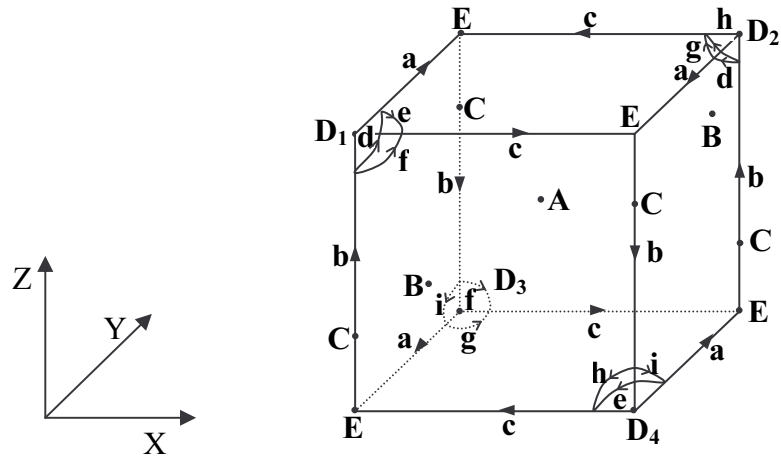

Figure 6: A 3-pseudomanifold

Here (Fig. 6) the neighborhoods of the points A (interior to the cube), B (interior to a face) and C (interior to an edge) behave essentially as in the former example, namely, they are 3-balls. (But notice the different locations of B and C, caused by our identification.) However, unlike the former example, here the eight vertices of the cube do not form one single point by the identification. Here they are grouped into two groups: the four verices D form one point and the four vertices E yield another point.

The four eighth of a 3-ball which together form the neighborhood of the point D (as well as those of E) do not fit together to form a 3- ball. In fact, their gluing cannot be done physically in the Euclidean 3-space. A careful check of the boundary of that neighborhood, that is, of linkD (composed of four sections of a 2-sphere) reveals (Fig. 7) that linkD is a projective plane $\mathbf{P}^2$. Similarly linkE too is $\mathbf{P}^2$.

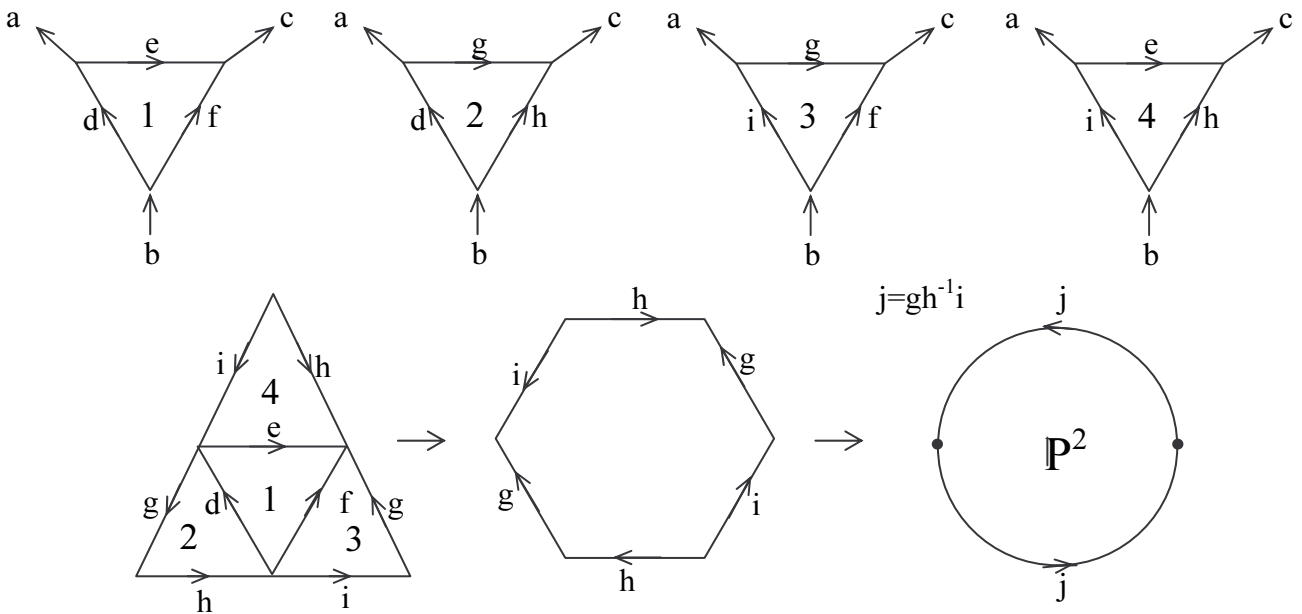

Figure 7: LinkD, a projective plane.

Thus the given identification yields from the 3-cube a 3-pseudomanifold with precisely two singular points. The fact that the number of singular points in this 3-pm, with links of odd Euler characteristics is even, is not a coincidence, as we will see later.

## 4. TRIANGULATIONS

In the following we wish to explain briefly and exemplify the concept of 3-pm and its singular points from a slightly different point of view.based on **triangulations**, The advantage of this approach is that it enables one to construct 3-pms with upriori given list of singularities almost at wish, and it enables the proof of the main theorem concerning 3- pms, stated in Section 5. . For a more detailed description one can refer to [2] and [5].

Consider a finite set of topological 3-simplices, namely tetrahedra whose edges and 2-faces (triangles) are not necessarily straight (such a tetrahedron with vertices w, x, y, z will be denoted wxyz and its 2-faces are the triangles wxy, wxz, wyz and xyz) such that every 2-face of each 3-simplex is also a 2-face of precisely one other 3-simplex in the set. We further assume that the 3-simplices and their faces form a **complex**, that is, the intersection of every two faces is a face (possibly empty) of both. (The 0-faces, 1-faces, 2-faces and 3-faces are the vertices, edges, triangles and tetrahedra , respectively.) Note that such a set cannot be realized in any bounded portion of the Euclidean 3-space. (However, it can be realized in higher dimensional space.)

In such a complex C, the neighborhood of every point interior to a tetrahedron is obviously a 3-ball. The same holds for each point interior to a 2-face (one half of the 3-ball being in one of the tetrahedra containing that 2-face, the other half in the other tetrahedron). If, in addition, the neighborhood of each vertex too is a 3-ball (which implies also that the neighborhood of each point interior to a 1-face of the complex C too is a 3-ball, see [8, p. 121 ] , [2, Th. 10]), then C (more precisely: the union of its tetrahedra ) is a 3-manifold. . Every 3-manifold can be endowed with such a triangulation [7 ].

Now, the requirement that the neighborhood of a vertex x in C is a 3-ball is equivalent to the requirement that the union of all the tetrahedra in C of which x is vertex (this union is called the **star** of x in C) is a 3-ball, and this in turn is equivalent to the requirement that the boundary of this union (named the **link** of x in C) is a 2-sphere. If the tetrahedra containing the vertex x are xabc, xdef. xghi,…then the link of x in C is the union of the triangles abc, def, ghi,…and it is easy to check whether or not this union is a 2-sphere.

As an example , consider the following complex $C_1$, given by means of the list of its tetrahedra (the vertices are denoted by numerals):

$C_1$: 1234, 1235, 1245, 1345, 2346, 2356, 2456, 3456.

Here the link of the vertex 2 is composed of the triangles 134, 135, 145, 346, 356 and 456. The union of these six triangles is easily seen to be a 2-sphere, as depicted in Fig. 8. (Recall that the outer triangle 145, which "closes" the 2-sphere, is one of the six.)

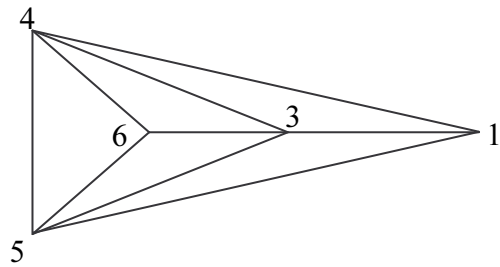

Figure 8: Link(2, $C_1$), a 2-sphere $\mathbf{S}^2$

Similarly, the link of each of the other five vertices too is easily checked to be a 2-sphere. Thus the conclusion follows that $C_1$ is a 3-manifold. (This 3-manifold is easily seen to be a 3-sphere, being the boundary of the 4-polytope obtained from the 4-tetrahedron 12345 by subdividing its face 2345 into four tetrahedra sharing a common vertex 6.)

Next consider the following complex $C_2$ composed of 27 tetrahedra (for the moment ignore the stars):

$C_2$:  1236, 1238, 1245, 1247, 1258, 1267, 1345, 1346, 1358, 1469, 1479, 1679, 2345*, 2346*, 2359*, 2389, 2467*, 2589, 3578*, 3579*, 3789, 4678*, 4689, 4789, 5678*, 5679, 5689.

This complex too satisfies all the above conditions and the link of each vertex is a 2-sphere. ( This complex is highly symmetric: The links of all the nine vertices are isomorphic as triangulated 2-spheres. Each is composed of twelve triangles. E.g., the link of the vertex 9 is as depicted in Fig. 9.

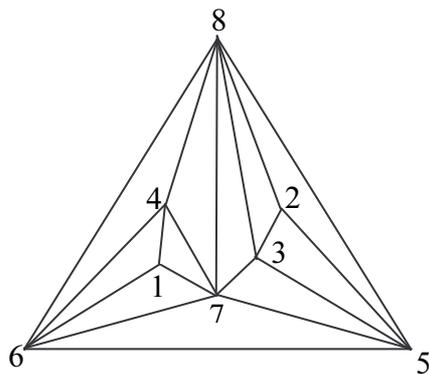

Figure 9: Link(9, $C_2$), a 2-sphere $\mathbf{S}^2$

(Recall that the outer triangle 568, which "closes" the 2-sphere, is one of the twelve.) Thus we conclude that $C_2$ is a 3-manifold. However, this 3-manifold is not a 3-sphere. It is a non-orientable 3-manifold (a concept that has not been defined here!). The union of the eight starred tetrahedra is a solid Klein bottle; its boundary is a "normal" Klein bottle. The complex $C_2$ is studied in detail in [3], named there $N^9_{51}$.

## 5. 3-PSEUDOMANIFOLDS

Now consider the following complex $C_3$:

$C_3$:  1245, 1235, 1356, 1346, 1467, 1457, 1267, 1237,
 1348, 1248, 1378, 1268, 1578, 1568, 2458,
 2358, 3568, 3468, 4678, 4578, 2678, 2378.

This complex too satisfies the condition that each 2-face is shared by precisely two tetrahedra. Also the link of each of the six vertices 2, 3,… 7 is a 2-sphere, but the link of each of the vertices 1, 8 is not a 2-sphere, it is a torus. Thus $C_3$ is not a 3-manifold. It is "almost" a 3-manifold, in the sense that the neighborhood of each of its (infinitely many) points but the two points 1 and 8 is a 3-ball. The neighborhood of each of the two points 1 and 8 is, so to speak, a "solid" torus, in the sense that its boundary is a regular 2-dimensional torus. Thus $C_3$ is a 3-pm with two singularities , and those singularities are tori.

The link of the vertex 1 in $C_3$ is depicted in Fig. 10.

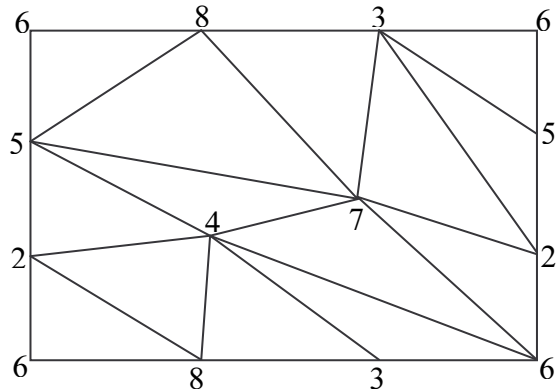

Figure 10:  Link(1, $C_3$),  a torus $\mathbf{T}^2$

As a last example, consider the following complex $C_4$:

$C_4$ :  1237, 1267,  1357, 2347, 1467, 1457, 3567, 3467
 2457, 2567, 1234, 1245, 1346, 1256, 1356

This complex too satisfies the condition that each 2-face is shared by precisely two tetrahedra. Also the link of each of the five vertices 2, 3,… 6 is a 2-sphere, but the link of each of the vertices 1, 7 is not a 2-sphere, it is a projective plane. Thus $C_4$ is 3-pm..   The link of the vertex 7 in $C_4$ is depicted in Fig. 11.

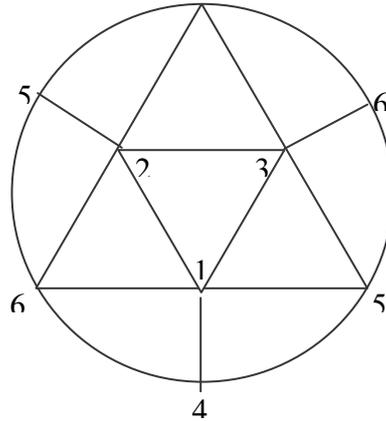

Figure 11: Link(7, $C_4$), a projective plane $\mathbf{P}^2$.

Now we can state the main theorem for 3-pms (see [2]):

**Theorem 1**: For every finite list of 2-manifolds (whether orientable or not, tori of any genus, projective planes, Klein bottles etc.), there exists a 3- pseudomanifold whose singularities are precisely the 2-manifolds in the list , if and only if the number of non-orientable members in the list that are of odd connectivity, is even.

## 6. GENERALIZED PSEUDOMANIFOLDS

The notion "pseudomanifold" can be generalized to include much richer types than presented so far, (see [5].)
Define a (combinatorial**) d-generalized pseuodomanifold** (abbrev. d-gpm) as a
 d-dimensional simplicial complex in which every (d-1)- face is contained in
exactly two facets (= d-faces). It follows than, that the link of every vertex in a d-gpm can be any (d-1)-gpm. Equivalently, a d-gpm can be defined inductively as a d-dimensional simplicial complex in which the link of every vertex is a (d-1)-gpm, where a one-dimensional gpm is one or more disjoint circles.
Thus a 2-gpm is basically a union of (originally) disjoint 2-manifolds or2-pseudomanifolds in which there is a finite set of finite sets of points that are identified to form a single point each, the link of this point being a disjoint union of circles.
Already the 2-gpms with only seven vertices include (beside the obvious 2-manifolds: sphere, torus and projective plane) two 2-spheres identified to each other in a single or in two points, and a 2-sphere identified to itself in a single point. In each of these cases the link of every singular point (= point of identification) is two disjoint circles.
 Each of those 2-gpm's can be (and is ) the link of some vertex in some 3-gpm.

D-gpms have been studied in [5]. Among the results obtained there is the following general theorem for the possible types of 3-gpms.

**Theorem 2** [5, p.51]**:** For every set $\sum$ of topological types of 2-gpms there is a 3-gpm in which the set of types of the links of the singular vertices is precisely $\sum$.

In a sense, the structure of this theorem resembles that of Theorem 1, but one should notice the basic difference, beside the fact that Theorem 1 deals with pms and Theorem 2 with gpms:: The list of 2-manifolds in Th. 1 may contain a certain 2-manifold M several times, and this is precisely the number of times that M will appear as a link in the promised 3-pm.. On the other hand, in Th. 2, $\sum$ is a set, not a list. As such, each type of 2-gpm appears there just once, and we have no control over how many times it appears as a link in the final 3-gpm.

CONCLUSION

Three-dimensional pseudomanifolds have been thoroughly investigated ([1], [2], [5]). We don't know of a similar investigation for higher-dimensional pseudomanifolds. In any case, we believe that the mathematical existence of pseudomanifolds of any dimension (established e.g. in [2, Th.14]) should draw attention to the possibility that the space-time in which we live is a pseudomanifold. Obviously, a similar remark can be made with generalized pseudomanifolds instead of pseudomanifolds.

REFERENCE

[1]. A. Altshuler , 3-pseudomanifolds with preassigned links, Trans. Amer. Math. Soc. **241** (1978) 213-237.
[2]. A. Altshuler and U. Brehm, The Topological Structure of 3-Pseudomanifolds, Israel J. Math. **39** (1981) 63-73.
[3]. A. Altshuler and L. Steinberg, Neighborly Combinatorial 3-Manifolds with 9 Vertices, Discrete Math. **8** (1974) 113-137.
[4]. A.T. Fomenko and T.L. Kunii, Topological Modeling for Visualization. Springer 1997.
[5]. G. Hanoch, On the problem of preassigning links for low-dimensional Pseudomanifolds, Ph.D. Thesis (Hebrew, English summary), Ben-Gurion univ. of the Negev, Beer-Sheva, Israel, 2001.
[6]. L.C. Kinsey, Topology of Surfaces. Springer 1993.
[7]. E.E. Moise, Affine structures in 3-manifolds, Ann. of Math. **56** (1952) 96 – 114.
[8]. W. P. Thurston, Three-Dimensional Geometry and Topology, Vol. 1, Princeton Univ. Press 1997.